\documentclass{article}
\usepackage{spconf,amsmath,graphicx}
\usepackage[utf8]{inputenc}
\usepackage{textcomp}
\usepackage[table]{xcolor}
\usepackage{tikz}
\usepackage{pgfplots}
\usepackage{acro}
\usepackage{relsize}
\usepackage{tikzscale}
\usepackage[super]{nth}

\usetikzlibrary{
  decorations.markings,
  shapes.geometric,
  arrows.meta,
  backgrounds,
  positioning,
  patterns,
  calc,
}

\acsetup{short-format=\scshape}

\DeclareAcronym{TTA} {short = TTA,  long = transport triggered architecture}
\DeclareAcronym{VLIW}{short = VLIW, long = very long instruction word}
\DeclareAcronym{FU}  {short = FU,   long = functional unit}
\DeclareAcronym{RF}  {short = RF,   long = register file}
\DeclareAcronym{LSU} {short = LSU,  long = load-store unit}
\DeclareAcronym{GCU} {short = GCU,  long = general control unit}
\DeclareAcronym{TCE} {short = TCE,  long = TTA-based Co-Design Environment}
\DeclareAcronym{TUT} {short = TUT,  long = Tampere University of Technology}
\DeclareAcronym{AG}  {short = AG,   long = address generator}
\DeclareAcronym{TFG} {short = TFG,  long = twiddle factor generator}
\DeclareAcronym{CADD}{short = CADD, long = complex adder}
\DeclareAcronym{FFT} {short = FFT,  long = fast Fourier transform}
\DeclareAcronym{LSB} {short = LSB,  long = least significant bit}
\DeclareAcronym{MSB} {short = MSB,  long = most significant bit}
\DeclareAcronym{LUT} {short = LUT,  long = lookup table}
\DeclareAcronym{ROM} {short = ROM,  long = read-only memory}
\DeclareAcronym{OFDM}{short = OFDM, long = orthogonal frequency division multiplexing}
\DeclareAcronym{ASIC}{short = ASIC, long = application specific integrated circuit}
\DeclareAcronym{FPGA}{short = FPGA, long = field-programmable gate array}
\DeclareAcronym{CGRA}{short = CGRA, long = coarse-grained reconfigurable array}
\DeclareAcronym{GPP} {short = GPP,  long = general purpose processor}
\DeclareAcronym{GPU} {short = GPU,  long = graphics processung unit}
\DeclareAcronym{PE}  {short = PE,   long = processing element}
\DeclareAcronym{DIT} {short = DIT,  long = decimation-in-time}
\DeclareAcronym{DIF} {short = DIF,  long = decimation-in-frequency}
\DeclareAcronym{SNR} {short = SNR,  long = signal-to-noise ratio}

\title{Low-Power Programmable Processor for Fast Fourier Transform Based on Transport Triggered Architecture}

\name{Jakub Žádník and Jarmo Takala}
\address{
  Faculty of Information Technology and Communication Sciences, Tampere University, Finland\\
  \texttt{\{jakub.zadnik, jarmo.takala\}@tuni.fi}\\
}

\begin{document}
\ninept

\maketitle

\begin{abstract}
This paper describes a low-power processor tailored for fast Fourier transform
  computations where transport triggering template is exploited.
The processor is software-programmable while retaining an energy-efficiency
  comparable to existing fixed-function implementations.
The power savings are achieved by compressing the computation kernel into one
  instruction word.
The word is stored in an instruction loop buffer, which is more power-efficient
  than regular instruction memory storage.
The processor supports all power-of-two FFT sizes from 64 to 16384 and given 1
  mJ of energy, it can compute 20916 transforms of size 1024.
\end{abstract}

\begin{keywords}
Fast Fourier Transform, Transport Triggered Architecture, Application-Specific
  Instruction-Set Processor
\end{keywords}

\section{Introduction}
\label{sec:intro}

\Ac{FFT} is one of the most widely used signal processing algorithms thanks to
its ability to represent a time-domain signal in a frequency domain.
For example, \ac{FFT} is used in \ac{OFDM} systems, which are employed in
wireless communication devices.
Due to the popularity of embedded and battery-powered systems, minimizing power
consumption is a major objective.

Regarding power consumption, \ac{ASIC} implementations are considered as more
efficient compared to reconfigurable hardware (such as \ac{FPGA}, \ac{CGRA}) or
\acp{GPP}.
However, \ac{ASIC}-based \ac{FFT} processors are mostly fixed-function and lack
programmability.
While the reconfigurable fabric offers more silicon reusability than \ac{ASIC},
their functionality can be only modified by a hardware design process similar
to \ac{ASIC}.
The goal of this work is to propose a software-programmable mixed radix-4/2
\ac{FFT} processor with an energy-efficiency comparable to fixed-function
\ac{ASIC} implementations.
Other software implementations of \ac{FFT} are implemented on either \ac{GPP}
\cite{khelifi15} or a \ac{GPU} \cite{lyu16}.
However, both of these approaches aim for the best performance and do not
provide sufficiently low-power solutions.

Fixed-function \ac{ASIC} \ac{FFT} processors can be divided into two categories
- pipelined and memory based.
Pipelined architectures (\cite{tran16}, \cite{garrido16}, \cite{garrido18})
rely on a cascade of \acp{PE} processing the input data stream.
The intermediate results are stored in a distributed memory system.
Due to a higher number of \acp{PE}, pipelined architectures consume more power
and occupy larger silicon area than memory based architectures.
However, they usually have higher throughput, which can lead to a high
energy-efficiency.

Memory based architectures (\cite{bass99}, \cite{huang16}) typically have one
\ac{PE} and data is processed in a sequential fashion.
They typically use only one or two global memory elements, thus a conflict-free
memory access has to be maintained.
The proposed architecture is memory based, using a single-port data memories as
it allows for a convenient software-programmable implementation.

The processor was designed using a \ac{TTA} template \cite{corporaal97}.
It improves a previous \ac{FFT} processor \cite{pitkanen11} by further
increasing its energy-efficiency.
Several optimizations were applied to allow compressing the computation kernel
into only one - repeatedly executed - instruction word that can be executed in
a more energy-efficient way.

\section{FFT Algorithm}
\label{sec:fft}

The proposed processor supports all the power-of-two \ac{FFT} sizes from $2^6$
to $2^{14}$.
A mixed radix-4/2 algorithm was used, following a \ac{DIT} approach
\cite{saidi94} as it
provides better \ac{SNR} compared to \ac{DIF} approach\cite{chang08}.
Otherwise, they share the same arithmetic complexity.
Radix-4 is used in a majority of the stages because is requires less operations
per \ac{FFT} than the radix-2 algorithm \cite{pitkanen14}.
At the same time, radix-4 butterfly operation requires only trivial operations.
Higher radices require more complicated operations.
However, using only radix-4 would restrict the processor to only power-of-four
\ac{FFT} sizes.
Therefore, in the last stage of the computation, radix-2 butterflies are used
for \ac{FFT} sizes which can not be computed using radix-4 algorithm (i.e. for
\ac{FFT} sizes $2^{k}$ where $k$ is odd).
The computation follows an in-place approach where output samples are written
back to the same memory locations from which the operands were read.
This allows to utilize only one memory module of the size equal to the computed
\ac{FFT} size.

\section{Transport Triggered Architecture}
\label{sec:tta}

\Ac{TTA} \cite{corporaal97} is a processor template, which exposes its internal
datapaths to a programmer.
Similarly to \ac{VLIW} \cite{fisher83}, it utilizes long instruction words and
instruction-level parallelism.
The difference is that \ac{TTA} gives a programmer the control over the data
flow.
It is possible to bypass accesses to \acp{RF} by feeding results from one
\ac{FU} directly to the input of another.
Register bypassing reduces the required \ac{RF} size and hardware complexity
leading to significant power savings \cite{jaaskelainen15}.

The data transports are defined by a \emph{move} instruction - the only
instruction of the \ac{TTA}'s instruction set.
\emph{Moving} data into a \emph{trigger} port of a \ac{FU} triggers the desired
operation.
\Acp{FU} can also have \emph{operand} ports for additional data that can be
loaded anytime without triggering the operation.
Memory access is performed by \ac{LSU} in a similar way as any other
instruction.
A control unit responsible for instruction fetching, decoding, and executing is
also implemented as one of the \acp{FU}.
Data moves are distributed over an interconnection network consisting of
several parallel buses.
The number of the parallel buses determines the maximum number of instructions
that can be executed in parallel, i.e., the maximum number of simultaneous data
moves.

\section{Proposed Processor Architecture}
\label{sec:arch}
\label{sec:proc}

\begin{figure}[!tb]
  \begin{minipage}[b]{1.0\linewidth}
    \centering
    \centerline{\includegraphics[width=8.5cm]{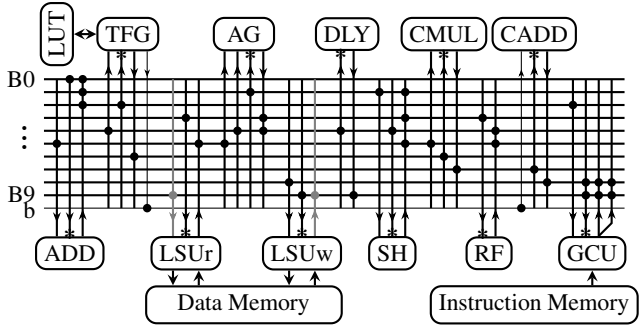}}
  \end{minipage}
  \caption{
    Architecture of the proposed TTA processor. \emph{TFG}: twiddle factor
    generator: \emph{AG}: address generator. \emph{DLY}: rotating register as a
    delay unit. \emph{CMUL}: complex multiplier. \emph{CADD}: complex adder.
    \emph{ADD}: adder. \emph{LSU}: load-store unit. \emph{SH}: shifter.
    \emph{RF}: 8x32 register file. \emph{GCU}: general control unit. Gray color
    denotes unused connections. Trigger ports are marked with `*'.
  }
  \label{fig:tta_arch}
\end{figure}

The proposed architecture is shown in Fig. \ref{fig:tta_arch}.
The architecture consists of ten 32-bit wide buses (B0--B9) and one 1-bit bus
(b), represented by horizontal lines.
\acp{FU} and one \ac{RF} are connected to the buses.
Vertical lines represent sockets, which connect input/output ports of \acp{FU}
to the interconnection network.
The connections are marked as dots.

Two \acp{LSU} are connected to a data memory system that behaves like a
dual-port memory.
In fact, two single-port memories are used and connected to the \acp{LSU} via
an added logic, which provides a conflict-free memory access.
The parallel memory system was chosen due to a lower power consumption of
single-port memories compared to multi-port memories \cite{pitkanen09}.

The parity of the address determines which one of the two single-port memories
is accessed.
In the case when both \acp{LSU} are trying to access an address with the same
parity (i.e. the same memory module), the processor is temporarily locked and
the accesses are resolved sequentially.
However, the conflict-free memory access is guaranteed for the \ac{FFT}
addressing scheme.

The streamlined instruction schedule (see Section \ref{sec:instr_sched})
implies generation of two parallel streams of addresses - read and write.
In order to guarantee a different parity for any two parallel addresses (thus
conflict-free memory access), a special scheduler module was put between the
\acp{LSU} and the parallel memory logic described in a previous paragraph.
The scheduler internally buffers and reschedules the \ac{LSU} data in a way
that always two parallel read addresses or two parallel write addresses are
loaded into the parallel memory logic.
Because the address generator preserves parity (see Section
\ref{sub:addr_gen}), the scheduler guarantees a conflict-free memory access.
The internal buffering is not recognized by a high-level compiler and,
therefore, the programming is only possible by low-level assembly.
However, it is possible to provide a software-exposed switch in a form of
another port of \ac{LSU} or a special \ac{FU} that toggles the scheduler on and
off, thus preserving a full compiler support for generic applications.

Loop buffer \cite{guzma10} - a critical component of the design - is
implemented as a part of the \ac{GCU}.
It is a small instruction memory cache used for storing frequently repeated
instruction words, e.g., loops.
Reading from a loop buffer consumes significantly less power than reading an
instruction directly from the instruction memory.

Each single-port data memory is composed of increasingly sized memory blocks
(32, 32, 64, 128, ..., 4096 - summing up to total 8192).
Based on the access address, only one block is selected at a time while the
other do not receive any control signals.
This significantly decreases dynamic power consumption when computing smaller
\ac{FFT} sizes.

The processor was designed using \ac{TCE} toolset developed at \ac{TUT}
\cite{tce17}.
\Ac{TCE} provides a comprehensive set of tools for designing \ac{TTA}
processors including a retargetable compiler and a hardware description of the
most common \acp{FU}.
For data and instruction memories, low power Cacti-P models were used
\cite{sheng11}.

\section{Special Function Units}
\label{sec:sfu}

This section describes special \acp{FU} developed specifically for this work.
All the other \acp{FU} were taken from \ac{TCE} component libraries.

Complex numbers are represented by two 16-bit fixed point numbers sharing one
32-bit data word.
The real part occupies \acp{LSB} of the data word while the imaginary part
takes the \acp{MSB}.

In order to prevent overflow, each addition is divided by two.
When summed up, the complex adder divides the result by four in case of radix-4
and by two in case of radix-2 butterfly.
The complex multiplier divides the result by two.

\subsection{Address Generator}
\label{sub:addr_gen}

The \ac{AG} is responsible for computing the memory addresses for butterfly
operands.
It is generated from a linear counter by a bit pair permutation following the
same pattern as the reference implementation \cite{pitkanen06}.
An example of an address generation for a 128-point and 256-point \ac{FFT} is
illustrated in Fig. \ref{fig:addr_gen}.
Each $b_i$ represents an $i$-th bit of a linear counter.
The `index' bits are sufficient to represent the index within one stage while
`stage' determines the current stage of the computation.
The position of the \ac{LSB} bit pair is determined by the `stage' part of the
linear counter.

The address generator preserves the parity of the linear counter.
Thus, any two consecutive addresses have a different parity and if fed in
parallel into the parallel memory logic (described in \ref{sec:proc}), a
conflict-free memory access is guaranteed.

\begin{figure}[!tb]
  \centering
  \includegraphics[width=0.7\linewidth]{addr_gen_128_256.tikz}
  \caption{
    Address generation from a linear counter for 128-point (above) and
    256-point (below) \ac{FFT}.
  }
  \label{fig:addr_gen}
\end{figure}

\subsection{Twiddle Factor Generator}

The generation of twiddle factors is based on a \ac{LUT} implemented as a
single-port synchronous \ac{ROM} of pre-computed values.
It follows the same approach as the one described in \cite{pitkanen07}.
The address for the \ac{LUT} \ac{ROM} is computed from the linear index by a
bit permutation and scaling based on the current \ac{FFT} size.
Only $N/8+1$ complex coefficients need to be stored in the \ac{LUT}
\cite{pitkanen07}.
All the remaining coefficients can be reconstructed by a trivial manipulation
(negating and swapping the real and imaginary parts) of the stored
coefficients.
Therefore, in order to support the maximum 16384-point \ac{FFT}, the \ac{LUT}
has to contain 2049 coefficients.
A side function of the \ac{TFG} \ac{FU} is determining whether the current
stage is radix-4 or radix-2.
This information is then used by the \ac{CADD}.

\begin{figure}[!b]
  \begin{center}
    \begin{tabular}{|l|l|r|}
      \hline
      rx2 & cnt & result \\
      \hline
      0 & 00 & \texttt{a + \ \ b + c + \ \ d} \\
      0 & 01 & \texttt{a -   i*b + c +   i*d} \\
      0 & 10 & \texttt{a - \ \ b + c - \ \ d} \\
      0 & 11 & \texttt{a +   i*b - c -   i*d} \\
      \hline
      1 & 00 & \texttt{a + b} \\
      1 & 01 & \texttt{a - b} \\
      1 & 10 & \texttt{c + d} \\
      1 & 11 & \texttt{c - d} \\
      \hline
    \end{tabular}
  \end{center}
  \caption{
    An operation performed by a complex adder based on the value of its `rx2'
    input and an internal counter (`cnt').
    Four operands (\texttt{a}, \texttt{b}, \texttt{c}, \texttt{d}) and rx2 are
    constant until the next reset of the counter.
    \texttt{i} denotes an imaginary unit.
  }
  \label{fig:cadd}
\end{figure}

\subsection{Complex Adder}

The \ac{CADD} performs a butterfly operation on four inputs.
Based on its `rx2' input, it performs either one radix-4 or two radix-2
butterflies.

Traditionally, the \ac{CADD} would be implemented as a four-input \ac{FU} with
the four inputs buffered in register files before feeding them in parallel into
the \ac{CADD}'s ports.
However, due to the single-instruction kernel requirement, the register file
buffering is not possible since the data can be moved only to a single
location.
Therefore, the proposed \ac{CADD} \ac{FU} has one serial data input port and
performs the buffering internally.
This makes the \ac{FU} unusable for high-level programming since this mode of
operation can not be recognized by a high-level compiler.

Figure \ref{fig:cadd} shows the \ac{CADD}'s results based on its `rx2' input.
The `cnt' column is an internal counter that increments each time a data sample
is loaded into the \ac{FU}'s \emph{trigger} port.
Both signals form an opcode selecting the operation of the complex adder.

\subsection{Complex Multiplier}

The complex multiplier performs generic complex multiplication of two operands.
The proposed implementation requires four multipliers and two adders.

\subsection{Rotating Register}

Rotating register is used to delay the address of a butterfly's input sample
for the in-place computation.
After the butterfly operation is complete, the output of the rotating register
is used as an address for the results to store them back to the memory.

\section{Instruction Schedule}
\label{sec:instr_sched}

\begin{table*}[t!]
  \centering
  \begin{tabular}{|l|c|c|c|c|c|c|c|c|c|c|}
    \hline
                      & type         & tech. & volt. & freq  & WL     & t        & power & FFT/mJ & FFT/mJ norm. & programmable \\
                      &              & (nm)  & (V)   & (MHz) & (bits) & ($\mu$s) & (mW)  &        &              &     \\ \hline
    \cite{garrido16}  & pipelined    & 65    & 1.10  & 50    & 16     & 21.5     & 17.60 & 2641   & 3196         & no  \\ \hline
    \rowcolor{gray!20}
    proposed          & memory based & 28    & 0.60  & 450   & 16     & 11.4     & 4.19  & 20916  & 3243         & yes \\ \hline
    \cite{shami18}    & memory based & 65    & 1.20  & 500   & 16     & 2.6      & 170.0 & 2287   & 3292         & no  \\ \hline
    \cite{pitkanen11} & memory based & 130   & 1.50  & 250   & 16     & 20.6     & 60.40 & 802    & 3609         & yes \\ \hline
    \cite{bass99}     & memory based & 600   & 3.30  & 173   & 20     & 30.0     & 845.0 & 39     & 6058         & no  \\ \hline
    \rowcolor{gray!20}
    proposed          & memory based & 65    & 1.00  & 450   & 16     & 11.4     & 12.21 & 7171   & 7171         & yes \\ \hline
    \cite{huang16}    & memory based & 90    & 1.00  & 160   & 16     & 2.6      & 29.00 & 13360  & 18498        & no  \\ \hline
    \cite{garrido18}  & pipelined    & 55    & 0.90  & 18    & 16     & 1.5      & 8.88  & 77131  & 52865        & no  \\ \hline
  \end{tabular}
  \caption{Comparison of various \ac{FFT} processor architectures (1024-point \ac{FFT}).}
  \label{tab:comparison}
\end{table*}

The computation of one radix-4 butterfly can be visualized with the aid of a
reservation table in Fig. \ref{fig:sched_bfly}.
Each column represents one clock cycle.
Buses are represented by rows and their names (on the right) correspond to the
ones shown in Fig. \ref{fig:tta_arch}.
Gray square denotes that an instruction, i.e., data transfer, is executed on
the bus during the clock cycle.
The instruction (data \emph{move}) transferred on each bus is shown on the
left.
The syntax respects the following pattern: \textsf{source.port $\rightarrow$
destination.port}.
Source and destination are \acp{FU}.
Port can be \textsf{t} (trigger), \textsf{o} (operand), \textsf{r} (result) and
\textsf{rx2} (output port of \ac{TFG} signalizing whether the butterfly is
radix-4 or radix-2).

\begin{figure}[!t]
  \begin{minipage}[b]{1.0\linewidth}
    \footnotesize
    \centering
    \centerline{\includegraphics[width=8.5cm]{one_butterfly.tikz}}
  \end{minipage}
  \caption{Bus reservation table of computing one radix-4 butterfly}
  \label{fig:sched_bfly}
\end{figure}

Full \ac{FFT} is computed by repeating the above pattern multiple times every
four clock cycles.
At \nth{13} clock cycle, the bus utilization reaches 100\% and the instruction
word becomes constant until no new samples need to be computed.
Thus, the execution can be separated into three stages: prologue (first 13
cycles), kernel (length depends on \ac{FFT} size) and epilogue (last 13
cycles).
The size of the prologue and epilogue is constant for all \ac{FFT} sizes.
Because the kernel consists of only one repeated instruction word, it can be
loaded into the loop buffer from where it can be fetched consuming minimal
power.

Apart from the prologue, kernel and epilogue, a setup code consisting of 6
instructions is present to distribute static parameters between \acp{FU}.
Thus, the size of the complete code is 33 ($6 + 13 + 1 + 13$) instructions.
The architecture uses 51-bit wide instruction words.

\section{Evaluation}
\label{sec:eval}

The processor was synthesized using Synopsys Design Compiler and
two IC technologies were used - a 28 nm FDSOI low-power technology and another
65 nm technology.

In order to be able to compare different technologies, the energy was
normalized according to the following formula \cite{bass99}, \cite{chen08}:
\begin{equation}
  E_n = E \frac
    { L_{ref} U^2_{ref} \left( \frac{1}{3} W^2_{ref} + \frac{2}{3} W_{ref} \right) }
    { L U^2 \left( \frac{1}{3} W^2                   + \frac{2}{3} W \right) },
\end{equation}
where $E_n$ is the normalized energy; $E$, $L$, $U$ and $W$ are parameters of
the proposed architecture (energy, technology size, voltage and word length,
respectively); the $_{ref}$ suffix marks the reference technology (65 nm, 1.0
V, 16 bits).

Table \ref{tab:comparison} compares the proposed architecture with selected
state-of-the-art solutions and traditional architectures.
The chosen focus point is a 1024-point \ac{FFT} as a mid-point between the
smallest and largest supported \ac{FFT} sizes.
The frequency 450 MHz is close to the maximum achievable frequency (500 MHz for
28nm/0.60V and 550 for 65nm/1.00V).
The maximum achievable frequency is 1150 MHz with 28nm/1.10V technology.

\section{Conclusion}
\label{sec:concl}

In this paper, a low-power software-programmable \ac{FFT} processor was
proposed, which is
 is based on a \ac{TTA} template.
The key contribution is reducing the computation kernel into only one repeated
instruction word and executing it from a loop buffer instead of fetching from
an instruction memory every clock cycle.
This reduces the power consumption of an instruction memory to a negligible
value.
In order to achieve the instruction word compression, an internal buffering was
introduced in case of a complex adder and a memory access, which renders them
unusable by a high level language compiler.
However, it is possible to provide a software-accessible switch to disable the
memory buffering for more generic applications.
Additional functionality can be introduced by adding other functional units.
Synthesis power evaluation performed at two different \ac{ASIC} technologies
(28 nm and 65 nm) shows that the processor can provide an energy efficiency
comparable with fixed-function \ac{ASIC} processors.

\section{Acknowledgments}
\label{sec:ack}

The authors thank the following sources of financial support: Tampere
University of Technology Graduate School, Business Finland (FiDiPro Program
funding decision 40142/14), and  ECSEL JU project FitOptiVis (project number
783162).

\vfill\pagebreak

\bibliographystyle{IEEEbib}
\bibliography{bibliography}

\end{document}